\documentclass{iopjournal}

\begin{document}

\articletype{Article type} 

\title{Half a century of the theory of synchronization}

\author{Yoshiki Kuramoto$^1$}

\affil{$^1$Graduate School of Science, Kyoto University, Kyoto 606-8502, Japan}



\email{kuramotoyoshiki@gmail.com}


\begin{abstract}
This review offers an overview of the development of the theory of coupled oscillators and synchronization over the past half century from a perspective of my own contributions. Among such contributions, the following three specific works will be focused on, serving as some key points to illustrate the field's evolution. They are the derivation of (1) a simplest partial differential equation exhibiting spatio-temporal chaos (Kuramoto-Sivashinsky equation), (2) a solvable mathematical model describing synchronization phase transition (Kuramoto model), and the discovery of (3) coexistence of coherence and incoherence in nonlocally coupled oscillators (chimera states). It is emphasized that all these works resulted from the phase reduction of the complex Ginzburg-Landau equation (or its variants), i.e. the equation derived with a coworker in 1974 from a certain reaction-diffusion model. A quick overview will be given on how the above three works influenced the subsequent development of the field of coupled oscillators and synchronization. Finally, a few comments will be made on how the methods of dynamical reduction, such as the center-manifold reduction and phase reduction, are crucial for exploring this field in depth. \par

This article is a largely faithful reproduction of the content presented in my award lecture.
\end{abstract}

\section{Introduction}

It has been about half century since my research on the dynamics of coupled oscillators was started. Over this period, the science of coupled oscillators has developed tremendously far beyond the level expected initially. In this review article, my own research career will be looked back, and from that perspective, some backgrounds of such remarkable growth of this field will be reflected. \par

My first work on coupled limit-cycle oscillators was a short paper written with Toshio Tsuzuki in 1974~\cite{1}, where we derived perturbatively an equation known as the complex Ginzburg-Landau equation (CGL) from a certain reaction-diffusion model. \par

In retrospect, it is now realized that most of my subsequent works of relative importance were rooted in this short paper. Among others, the following three works will be focused on in this article. The first is the derivation of the equation called the Kuramoto-Sivashinsky equation~\cite{2}. The second work is the proposal of a mathematical model known as the Kuramoto model~\cite{3}, and the third work is a theoretical finding of a curious pattern of synchronization, which was the first example of the so-called chimera states~\cite{4}. Indeed, all these works resulted from CGL or its variants by applying a method called the phase reduction, i.e. a method of approximately describing limit-cycle oscillations only with phase variables. More details on how these works were produced and how they have given impact on the related fields will be developed in the subsequent sections. In the rest of this section, what motivated me to do the CGL work will briefly be described. \par

One thing which inspired me was that I came across an oscillatory reaction-diffusion model called the Brusselator~\cite{5}, a model invented by the Prigogine group of Brussels. Another source of my motivation was that I learned the self-organized wave patterns of the oscillatory Belousov-Zhabotinsky reaction~\cite{6} which was a topic of great popularity in those days. \par

While vaguely thinking about how to interrelate between the Brusselator and the wave patterns of the Belousov-Zhabotinsky reaction, one idea came to me. This idea which seemed rather promising was to focus on the situation near the onset of oscillation, namely the Hopf bifurcation of the Brusselator. It seemed quite probable that the diffusion-coupled Brusselator would then become much simpler and hopefully this would lead to a qualitative understanding of the observed wave patterns. \par

Fortunately, this idea was not wrong, and given shape by applying the perturbative method of Newell and Whitehead~\cite{7} who developed it for deriving a small-amplitude envelope equation near the onset of the Rayleigh-B\'enard convection. The reduced equation we obtained was CGL. The same equation can of course be derived also from a much more general reaction-diffusion equations by reduction~\cite{8,9}, and this type of reduction method is generally called the center-manifold reduction. \par

Somewhat related to this fact, it would be meaningful here to point out the dual nature of dynamical reduction in general, including the center-manifold reduction and phase reduction. Here the term ``dual nature'' is used to mean that formally a dynamical reduction is an approximate method but physically it goes far beyond a mere approximation. Indeed, much information may be lost by reduction, but the reward gained will be far richer. This is because the reduction removes so many restrictions imposed on the original specific model, and largely broadens our perspective. \par

To illustrate such potentiality of dynamical reduction, a simplest phase reduction of CGL will now be touched upon. The result of such reduction is a nonlinear phase diffusion equation with a constant frequency term. The reduction is valid when the spatial variation of the complex field is of sufficiently long scale. If necessary, the constant term can be dropped by working with a co-moving frame of reference. In one space dimension, and introducing Gaussian white noise, the equation becomes the famous Kardar-Parisi-Zhang (KPZ) equation of stochastic surface growth~\cite{10} which opened up a new research field of KPZ universality~\cite{11}. If our nonlinear phase equation is expressed using the spatial derivative of phase, the result is the Burgers equation~\cite{12}. There are significant amount of research works on this equation in connection with fluid dynamics and traffic flow~\cite{13,14}. In collaboration with Tomoji Yamada, our nonlinear phase diffusion equation was also studied in two dimensions, by introducing an additional term of localized heterogeneity. This led a successful explanation of the formation of expanding concentric wave patterns and their interaction similar to those observed in the Belousov-Zhabotinsky reaction~\cite{15}. These examples imply that the phase reduction of CGL is not a mere approximation of CGL, in much the same way that CGL itself is not a mere approximation of the Brusselator. 

\section{Kuramoto-Sivashinsky equation}

We now proceed to the partial differential equation called the Kuramoto-Sivashinsky (KS) equation. This equation was derived in 1976 as a phase reduction of CGL. As is well known, CGL with sufficient spatial extension admits a family of plane wave solutions over the whole range of wavelengths. However, they are not always stable. Generally, longer wavelength implies higher stability. But even the infinite wavelength plane wave, that is, spatially uniform oscillation, can become unstable. This is commonly called the Benjamin-Feir instability; it occurs for a certain range of parameter values of CGL. The instability condition itself had already been known before the derivation of the KS equation~\cite{16}, although there had been no argument as to what would happen right after the instability. The answer is spatio-temporal chaos governed by the KS equation~\cite{17,18}. \par

The KS equation can in principle be obtained in the following way~\cite{2}. Impose the hypothesis that the modulus of the complex field of CGL is given by a functional of the phase, or a function of various space derivatives of phase. Then, make an expansion of the modulus in powers of all its arguments. After substituting this expansion form into CGL, the expansion coefficients are determined. Then, out of various spatial derivatives of phase, single out the most dominant terms assuming that the instability is sufficiently weak. In doing this, a scale-invariance argument would be helpful. The resulting equation is the KS equation. After rescaling, the equation becomes parameter free. \par

Numerical simulation of this parameter-free equation for sufficiently large one-dimensional system revealed that the solution behaved chaotically in space and time. Such behavior is called the phase turbulence. Next year, Sivashinsky found the same equation independently~\cite{19,20} as an equation describing weakly unstable flame front of combustion. KS-like equations also appear in other physical contexts such as viscous liquid flowing down on a vertical wall~\cite{21}, epitaxial growth of solid surface by sputtering~\cite{22}, and instability of trapped ion mode in plasmas~\cite{23,24}. \par

As is widely known, the late 1970s was the period when the science of chaotic dynamical systems was experiencing a rapid surge. The major target of theoretical study there was ordinary differential equations and discrete maps exhibiting low-dimensional chaos. On the other hand, the phase turbulence we found was a chaotic solution of a partial differential equation, and not ordinary differential equation. A natural question was then raised from mathematics side. The question was ``Are ordinary differential equation (ODE) chaos and partial differential equation (PDE) chaos equivalent?'' This was answered affirmatively at least for the one-dimensional KS equation with finite system length. To be specific, the existence of the so-called inertial manifold was proved~\cite{25}. Inertial manifold is a finite-dimensional globally attracting manifold in which all infinite-time solutions come to be included. Consequently, the dynamics which is taking place within this finite-dimensional manifold becomes mathematically equivalent to ODE dynamics. This shows nothing more than the equivalence between ODE chaos and PDE chaos. \par

There was another issue raised from physics side. This was how a coherent structure changes to turbulence through successive bifurcations. Indeed, the scenario leading to turbulence had been one major issue of fluid dynamics since the classical theory of Landau~\cite{26} who imagined developed turbulence as a high-dimensional torus. However, the case of the Rayleigh-B\'enard convection, for instance, clearly contradicted this classical picture, because the existence of low-dimensional chaotic attractor was experimentally confirmed there~\cite{27}. Still the subsequent steps leading to developed turbulence remained unclear. In order to gain some insight into this problem, the KS equation was employed as a simplest analogue to the Navier-Stokes equation. Fortunately, the parameter-free one-dimensional KS equation includes only one control parameter, which is the system length $L$, and this quantity plays an analogous role to the aspect ratio of the Rayleigh-B\'enard convection. Extensive numerical study of the KS equation revealed~\cite{28} the detailed process of successive appearance of higher and higher dimensional chaotic attractors with increasing $L$, eventually leading to extended spatio-temporal chaos. \par

It was also found that the asymptotic statistical properties of the solution of the KS equation as the system length goes to infinity were extremely simple. For instance, the number of positive Lyapunov exponents as well as the attractor dimension defined by the Kaplan-Yorke dimension increased linearly with $L$~\cite{29,30}. This proved the extensivity of the system similar to that shared by a macroscopic body at thermal equilibrium. \par

It seems that most of the works on the KS equation in the past have made use of its simplicity as a minimal partial differential equation displaying spatio-temporal chaos. It seems, however, that there is another point of significance of this equation. That is, it represents a concise mathematical manifestation of the fact that diffusion-coupled fields of oscillators can easily display turbulence. This fact had not clearly been recognized before the appearance of the KS equation. Over time, however, numerous works appeared on the turbulent behavior of CGL~\cite{31,32} under the terms such as amplitude turbulence, spiral wave turbulence, and spatio-temporal intermittency.
 
\section{Kuramoto model}

The phase oscillator model called the Kuramoto model also has a close connection with CGL. My proposal of this model was inspired by Arthur Winfree's seminal paper of 1967 entitled ``Biological Rhythms and the Behavior of Populations of Coupled Oscillators''~\cite{33}. \par

Winfree's theory gave me a strong impression in the following two ways. The first is that he boldly described a limit-cycle oscillator only with one variable, that is phase, and the pair coupling with two phases. In this way, he successfully created a simple mathematical model for an infinitely large assembly of coupled oscillators. Although it is now commonly known that the phase description of coupled oscillators is applicable whenever the mutual coupling is sufficiently weak~\cite{9,34}, Winfree was the first to apply the same idea to collective dynamics. \par

The other point which Winfree's paper impressed me was that, under the assumption of global coupling, he claimed the existence of a critical condition for the onset of collective oscillation. Unlike macroscopic bodies composed of inactive elements such as atoms and spins, here the components are dynamically active. Having had some experience in the statistical mechanics of the second order phase transition, such a new type of phase transition was a fresh surprise to me. Winfree's phase transition is also unique in that the randomness comes from the heterogeneity in the natural frequencies of the oscillators and not random noise. Therefore, his model represents a deterministic dynamical system of infinitely high dimension. This point also struck me as quite intriguing. \par

Unfortunately, however, Winfree's argument on the existence of a phase transition was not convincing enough. This moved me to try to modify his model into a mathematically tractable model, preserving the original form as much as possible. The only modification needed was his coupling term which was given by a product of two functions representing influence and sensitivity. At this very point, my preceding work on CGL helped me greatly. The idea which came up with was that if a spatially discretized version of CGL is considered, then the phase reduction of the discretized diffusive coupling would become a simple trigonometric function depending only on the phase difference. Even a cosine component does not appear with vanishing imaginary part of some coefficients of the discretized CGL. Furthermore, it was assumed that the system size is infinitely large, and that the distribution of the natural frequencies is given by a unimodal symmetric function. The model obtained in this way eventually came to be called the Kuramoto model~\cite{35}. \par

Thanks to the global coupling and sinusoidal coupling function, this model was handled analytically by applying the elementary idea of the mean field theory similar to that of thermodynamics phase transitions. Actually, this was achieved by introducing a complex order parameter whose amplitude gives a measure of the strength of macroscopic oscillation. Although some nontrivial problems peculiar to the ensemble of heterogeneous dynamical elements were faced with, they were cleared by imposing a certain hypothesis, and a self-consistency equation for the order parameter was obtained. Its solution clearly showed the existence of a critical coupling strength above which a solution of collective oscillation bifurcated from a constant non-oscillatory solution. It seemed quite natural to expect that across the critical point an exchange of stability occurs between these solutions. Yet I could not find any mathematical evidence to support this seemingly obvious fact. In fact, this stability problem was to become one major issue in later years as will be touch upon in a later paragraph. \par

In any case, finishing this work brought me some sense of satisfaction, but nothing more. Thus, what I did was only to present a short talk on this model at a certain international symposium held in Kyoto, and submit a short article for the Proceedings~\cite{3}. Naturally, this article remained unnoticed for the subsequent several years. Probably, the first person who discovered it with great enthusiasm was none other than Winfree himself. It seems that he came across my article in his effort of searching for all relevant literatures for his book in progress entitled ``The Geometry of Biological Time''~\cite{36} which was published in 1980. Encouraged by this, four years later, some details of my model were included in my own book ``Chemical Oscillations, Waves, and Turbulence''~\cite{9}. \par

We now come back to the stability problem of the collective states~\cite{37}. Indeed, their stability was not obvious at all. This is because unlike thermodynamic phase transition, our transition has no potential or Lyapunov function like free energy. The conventional bifurcation theory of dynamical systems is also inapplicable. To put the nontrivial nature of the problem shortly, when we try to examine the stability of the incoherent branch of the solution using a standard method, we are faced with a peculiar spectrum such that the spectrum is distributed continuously on the imaginary axis. Formally this implies neutral stability of the incoherent state. However, this seemed contradictory to the asymptotic stability of the corresponding macroscopic state confirmed by numerical simulation. This theoretical dilemma aroused considerable mathematical curiosity. The clue to the solution was to correctly formulate the phase mixing mechanism analogous to the Landau damping of collisionless plasmas~\cite{38}. Thanks to the efforts of a number of theorists including Strogatz, Mirollo and Crawford, this issue, including the post-critical dynamics, was basically settled by the end of the last century~\cite{37,39}. Still, it took another 15 years or more until a fully satisfactory mathematical solution was finally given by Hayato Chiba~\cite{40}. \par

Setting the stability issue aside, since the beginning of this century, the research on the dynamics of coupled oscillators has been given a fresh momentum. The most widely studied model was sinusoidally coupled phase oscillators, often called the Kuramoto oscillators. There were at least two factors which pushed forward such a rapid advancement of this research field. One is the innovative theory of Edward Ott and Thomas Antonsen~\cite{41} which appeared in 2008, and the other is the appearance of mathematical models of complex networks. Ott-Antonsen's theory concerns large assemblies of Kuramoto-type phase oscillators with global coupling and randomly distributed natural frequencies. What is remarkable about their theory is that it achieves a drastic reduction of the dynamics of a large assembly to macroscopic dynamics of extremely low dimension. In particular, for a certain class of frequency distribution, an evolution equation for the order parameter can be written down in a closed form. The impact of the Ott-Antonsen theory was so great, broadening dramatically the class of problems that can be handled analytically. \par

The science of complex networks also expanded largely the scope of the dynamics of coupled oscillators. Around the turn of the century, two mathematical models of complex networks were proposed in succession. They are the small-world network of Watts and Strogatz~\cite{42}, and the other is the scale-free network of Barab\'asi and Albert~\cite{43}. \par

For a long time before the appearance of these network models, over nearly a quarter century, the study of large systems of coupled oscillators had been limited exclusively to the systems of regular connection topology, classified only according to the coupling range, like global coupling, local coupling, and their intermediate. However, with the sudden upsurge of the field of complex networks, the situation changed dramatically. The central issue of concern there was how the relation is between the network topology and the dynamics~\cite{44}. A vast number of research works appeared on this issue. They include the study of how the synchronizability is changed with the change of small-world-ness of the network, and, as for scale-free network, how the critical condition for the onset of global synchrony depends on the degree distribution and other network characteristics such as the degree of clustering. \par

In parallel with such generalizations from regular to complex topologies, my original phase oscillator model has been generalized in various ways. A few of them are the following. As for the connection scheme, a notable recent generalization is to introduce many-body interaction, or the so-called higher-order interaction, in place of the traditional pair coupling~\cite{45,46}. The research on this class of systems seems to be rapidly growing to form an important branch of science of coupled oscillators and other many-body systems in the real world. As a generalization of the coupling function, the importance of time delay in some practical situations was suggested~\cite{47}. The intrinsic dynamics of the individual oscillator was also generalized to include the effect of inertia~\cite{48}. Another interesting generalization is to extend the phase to a multi-dimensional variable defined on a $D$-dimensional sphere~\cite{49}. These generalizations and those not listed here were partly required from practical purposes. For instance, the effect of inertia was thought necessary for the application to power-grid dynamics~\cite{50}, and extending the phase to a multi-dimensional variable was intended for the application to such problems as opinion dynamics~\cite{51} and even machine learning~\cite{52}. As an important consequence of the study of such generalized phase oscillator models, it has become clear that synchronization phase transition is not always continuous, but discontinuous or explosive synchronization occurs rather easily~\cite{53}. Regarding practical significance of the explosiveness and the resulting multi-stability, we expect further investigation. All in all, the scope of the application of the phase oscillator models has expanded across a remarkably wide range of scientific disciplines, and it is already far beyond my ability to follow.

\section{Chimera states}

We now move to the issue of chimera states. A chimera state generally refers to a state where coherent and incoherent parts coexist in a system composed of identical oscillators coupled homogeneously. The first example of chimera states was found in 2002~\cite{4} in my joint work with Dorjsuren Battogtokh while studying nonlocally coupled oscillators. Preceding this work, going back seven years or so, my interest had been directed toward nonlocally coupled oscillators, expecting that some new type of dynamics might emerge. From a practical point of view also, nonlocal coupling seemed quite meaningful. For example, consider a mathematical model such that the coupling of uniformly distributed oscillators is mediated by some rapidly diffusing and decaying substance which is being produced from the oscillators themselves. Then, if we eliminate adiabatically this mediator, the effective oscillator coupling becomes nonlocal. Situations similar to this seemed likely to be realized in biological settings. It was also noticed that the mean field theory could be used for nonlocal coupling if we assume that the oscillators are distributed densely enough within the coupling radius. \par

As a simple mathematical model, a nonlocal version of CGL in one dimension with exponential coupling kernel~\cite{54} was considered. My first question was what would happen if the Benjamin-Feir instability occurred in this system causing turbulence. As long as the turbulence remained sufficiently weak, the phase fluctuations generated would be limited to a wavelength regime much longer than the coupling radius. Then, the nonlocal coupling would practically be the same as local coupling, and nothing new would happen. But how about if the turbulence became stronger and stronger, generating fluctuations with shorter and shorter wavelengths violating finally the coupling range? Well within the coupling range, there is no more characteristic length scale. Consequently, the fluctuations occurred in this scale-free domain were expected to obey self-similar statistics. Numerical simulation confirmed this~\cite{55}, and its theoretical explanation was also given. This was a joint work with Hiroya Nakao, a graduate student at that time, and also with Battogtokh, who joined us a little later~\cite{56,57}. \par

The phase reduction of the same model was also studied. But, as long as the nature of turbulent fluctuations is concerned, no particularly interesting results were found. In contrast, under a non-turbulent condition, something peculiar occurred which later came to be called a chimera state. \par

This was numerically found by Battogtokh sometime after he left Kyoto and moved to the University of Georgia. The data he sent me showed that nonlocally coupled identical phase oscillators distributed on a ring split into two regions. In one region, the oscillators were completely synchronous, while in the other region, their phases were randomly scattered. In the latter incoherent region, the drift velocities of the phases were distributed but they formed a systematic curve with a single hump. It seemed that such symmetry-breaking behavior could be accounted for in the framework of the mean field theory. For this purpose, a local order parameter was introduced as a space-dependent generalization of the global order parameter, the former being defined by a weighted local average over the coupling range instead of global average. By imposing a hypothesis that the local order parameter is steadily oscillating, a functional self-consistency equation was obtained for this quantity. Numerical analysis of this equation revealed that other than a trivial constant solution of full amplitude corresponding to global synchrony, a spatially nonuniform solution existed. The reason why the latter corresponds to the coexistence of coherence and incoherence is easy to see. According to the mean field picture, the oscillators can be regarded as mutually independent except that they are under the forcing from the mean field. This implies that there should be a critical amplitude in the local order parameter below which the individual oscillators fail to synchronize with the weak forcing of the mean field, implying incoherence, while above which they synchronize with the strong forcing, implying coherence. In consistent with such reasoning, our simulation results were reproduced by the theory~\cite{4}. \par

Our theory attracted the attention of Abrams and Strogatz. And two years later, they presented a more detailed analytical study of the same model as ours except one point that our exponential coupling kernel was replaced with cosine kernel~\cite{58}. In the same paper, they coined the term ``chimera states'' for such coexistence states of coherence and incoherence. This evocative term certainly contributed greatly to the subsequent development of the research field of complex synchronization patterns. Abrams and Strogatz have also made other important contributions about chimera states, but in what follows another contribution on our side will be introduced. \par

This was based on the idea that if a steadily rotating spiral pattern appeared in a nonlocally coupled oscillator field, then this pattern would necessarily include a local group of incoherent oscillators other than a coherent group. The reason is the following. Unlike the case of locally coupled oscillators, the oscillator sitting right at the center of rotation need not have vanishing amplitude. What must vanish at the center is the local order parameter. This implies that the local order parameter in some region around the center of rotation will have small amplitude, and that will be too weak to entrain a certain group of oscillators distributed there, whereas well apart from the center, the local order parameter will be large enough to entrain. Consequently, the pattern will split into two regions, that is, incoherent inner region and coherent outer region. This was actually proved by numerical simulation for the FitzHugh-Nagumo oscillators and also for phase oscillators each with nonlocal coupling. The corresponding mean field theory was also formulated. This work was done by Shin-ichiro Shima and myself in 2004~\cite{59}. \par

Chimera states were also realized in experiments, for example, in opto-electronic systems~\cite{60}, photo-sensitive Belousov-Zhabotinsky reaction~\cite{61}, and mechanical systems such as coupled metronomes and pendulums~\cite{62}. Theoretically, chimera states were found in a variety of topological settings, and many variants of chimera states were proposed under different naming~\cite{63}. 

\section{Conclusion}
 
In concluding this article, a few additional comments on dynamical reduction will be made. We have frequently referred to the two reduction schemes. One is the center-manifold reduction by which CGL was derived, and the other is the phase reduction. In the three specific works discussed in this article, these reduction methods were applied in conjunction. That is, CGL and its variants, which are themselves reduced equations, were further reduced to phase equations. As is well known, the phase reduction itself can of course be applied independently to more general class of oscillators~\cite{64}, not limited to the CGL-like oscillators with good symmetry, and in fact this method alone has produced many fruitful results. \par

In my view, the significance of dynamical reduction is twofold. First, as anyone may agree, it cuts off many details which seem inessential for the basic understanding of the system under study, thus making the analysis far easier. Even when we have to come back to the original complex model to know more details, the analysis of the reduced model may provide a useful guidance. The second point of significance of dynamical reduction is more conceptual than practical, where we are more concerned with a broad class of systems rather than individual specific systems. Indeed, reduction reveals something universal lurking behind apparently different systems and circumstances. Many scientists in nonlinear dynamics and statistical physics have taken attitude along this line of thought. Thanks to this, our modern view of this complex world has been largely broadened and integrated. As for my own phase oscillator model, it may have served as a small piece of such universality. It is an extremely boiled down model as an inevitable result of double reductions, and therefore it could only capture the crudest level of universality. Still, for its extreme simplicity, it leaves much room for various modifications and generalizations which had in fact been achieved over the past half century. The broad class of oscillator models thus produced will be of a great help for further understanding and controlling what is going on in the real world full of oscillations and synchronization.









\begin{thebibliography}{99}

\bibitem{1}
Kuramoto Y and Tsuzuki T, {\it Reductive perturbation approach to chemical instabilities}, 1974 {\it Prog. Theor. Phys.} {\bf 52} 1399

\bibitem{2}
Kuramoto Y and Tsuzuki T, {\it Persistent propagation of concentration waves in dissipative media far from thermal equilibrium}, 1976 {\it Prog. Theor. Phys.} {\bf 55} 356

\bibitem{3}
Kuramoto Y, {\it Self-entrainment of a population of coupled non-linear oscillators}, 1975 In {\it International symposium on mathematical problems in theoretical physics}, {\it Lecture Notes Phys.} {\bf 39} 420

\bibitem{4}
Kuramoto Y and Battogtokh D, {\it Coexistence of coherence and incoherence in nonlocally coupled phase oscillators}, 2002 {\it Nonlin. Phenom. Complex Sci.} {\bf 5} 380

\bibitem{5}
Glansdorff P and Prigogine I, {\it Thermodynamic Theory of Structure, Stability, and Fluctuations} (Wiley 1971)

\bibitem{6}
Zaikin AN and Zhabotinsky AM, {\it Concentration wave propagation in two-dimensional liquid-phase self-oscillating systems}, 1970 {\it Nature} {\bf 225} 535

\bibitem{7}
Newell AC and Whitehead JA, {\it Finite bandwidth, finite amplitude convection}, 1969 {\it J. Fluid Mech.} {\bf 38} 279

\bibitem{8}
Kuramoto Y and Tsuzuki T, {\it On the formation of dissipative structures in reaction-diffusion systems: reductive perturbation approach}, 1975 {\it Prog. Theor. Phys.} {\bf 54} 687

\bibitem{9}
Kuramoto Y, {\it Chemical Oscillations, Waves, and Turbulence} (Springer 1984)

\bibitem{10}
Kardar M, Parisi G and Zhang Y-C, {\it Dynamic scaling of growing interface}, 1986 {\it Phys. Rev. Lett.} {\bf 56} 889

\bibitem{11}
Takeuchi KA, {\it An appetizer to modern developments on Kardar-Parisi-Zhang universality class}, 2018 {\it Physica A} {\bf 504} 77

\bibitem{12}
Burgers JM, {\it The Nonlinear Diffusion Equation: Asymptotic Solutions and Statistical Physics} (Reidel 1974)

\bibitem{13}
Beck J and Khanin K, {\it Burgers turbulence}, 2007 {\it Phys. Rep.} {\bf 447} 1

\bibitem{14}
Lighthill MJ and Whitham GB, {\it On kinetic waves II: theory of traffic flow on long crowded roads}, 1955 {\it Proc. Roy. Soc. A} {\bf 229} 317

\bibitem{15}
Kuramoto Y and Yamada T, {\it Pattern formation in oscillatory chemical reactions}, 1976 {\it Prog. Theor. Phys.} {\bf 56} 724

\bibitem{16}
Lange CG and Newell AC, {\it A stability criterion for envelope equations}, 1974 {\it SIAM J. Appl. Math.} {\bf 27} 441

\bibitem{17}
Yamada T and Kuramoto Y, {\it A reduced model showing chemical turbulence}, 1976 {\it Prog. Theor. Phys.} {\bf 56} 681

\bibitem{18}
Kuramoto Y, {\it Diffusion-induced chaos in reaction systems}, 1978 {\it Prog. Theor. Phys. Suppl.} {\bf 64} 346

\bibitem{19}
Sivashinsky GI, {\it Nonlinear analysis of hydrodynamic instability in laminar flame I: derivation of basic equations}, 1977 {\it Acta Astronautica} {\bf 4} 1177

\bibitem{20}
Sivashinsky GI, {\it On self-turbulization of a laminar flame}, 1979 {\it Acta Astronautica} {\bf 6} 569

\bibitem{21}
Sivashinsky GI and Michelson DM, {\it On irregular wavy flow of a liquid film down a vertical plane}, 1980 {\it Prog. Theor. Phys.} {\bf 63} 2112

\bibitem{22}
Emel'yanov VI, {\it The Kuramoto-Sivashinsky equation for the defect-formation instability of a surface-stress}, 2009 {\it Laser Phys.} {\bf 19} 538

\bibitem{23}
LaQuey RE, Mahajan SM, Rutherford PH and Tang WM, {\it Nonlinear saturation of the trapped-ion mode}, 1975 {\it Phys. Rev. Lett.} {\bf 34} 391

\bibitem{24}
Cohen BI, Krommes JA, Tang M and Rosenbluth MN, {\it Nonlinear saturation of the dissipative trapped-ion mode by mode coupling}, 1976 {\it Nuclear Fusion} {\bf 16} 971

\bibitem{25}
Foias C, Nicolaenko B, Sell GR and Temam R, {\it Inertial manifold for the Kuramoto-Sivashinsky equation and estimate of their lowest dimension}, 1988 {\it J. de Math. Pures et Appl.} {\bf 67} 197

\bibitem{26}
Landau LD, {\it On the problem of turbulence}, 1944 {\it C. R. (Doklady) Acad. Sci. USSR} {\bf 44} 311

\bibitem{27}
Libchaber A, Maurer J and He X, {\it Experimental study of the transition to chaos in Rayleigh-B\'enard convection}, 1980 {\it J. de Phys.} {\bf 41} 515

\bibitem{28}
Hyman JM and Nicolaenko B, {\it The Kuramoto-Sivashinsky equation: a bridge between PDE's and dynamical systems}, 1986 {\it Physica D} {\bf 18} 113

\bibitem{29}
Manneville P, {\it Lyapunov exponents for the Kuramoto-Sivashinsky model}, 1984 In {\it Macroscopic modeling of turbulent flows and fluid mixtures}, {Lecture Notes Phys.} {\bf 230} 319

\bibitem{30}
Manneville P, {\it Dissipative Structures and Weak Turbulence} (Academic Press 1990)

\bibitem{31}
Egolf DA and Greenside HS, {\it Characterization of the transition from defect to phase turbulence}, 1995 {\it Phys. Rev. Lett.} {\bf 74} 1751

\bibitem{32}
Shraiman BI, Pumir A, van Saarloos, Hohenberg PC, Chat\'e H and Holen M, {\it Spatio-temporal chaos in the one-dimensional complex Ginzburg-Landau equation}, 1992 {\it Physica D} {\bf 57} 2451

\bibitem{33}
Winfree AT, {\it Biological rhythms and the behavior of populations of coupled oscillators}, 1967 {\it J. Theor. Biol.} {\bf 16} 15

\bibitem{34}
Hoppenstead FG and Izhikevich EM, {\it Weakly Connected Neural Networks} (Springer 1997)

\bibitem{35}
Acebr\'on JA, Bonilla LL, P\'erez Vicente CJ, Ritort F and Spigler SH, {\it The Kuramoto model: a simple paradigm for synchronization phenomena}, 2005 {\it Rev. Mod. Phys.} {\bf 77} 137

\bibitem{36}
Winfree AT, {\it The Geometry of Biological Time} (Springer 1980)

\bibitem{37}
Strogatz SH, {\it From Kuramoto to Crawford: exploring the onset of synchronization in populations of coupled oscillators}, 2000 {\it Physica D} {\bf 143} 1

\bibitem{38}
Strogatz SH, Mirollo RE and Matthews PC, {\it Coupled nonlinear oscillators below the synchronization threshold: relaxation by generalized Landau damping}, 1992 {\it Phys. Rev. Lett.} {\bf 68} 2730

\bibitem{39}
Crawford JD and Davies KTR, {\it Synchronization of globally coupled phase oscillators: singularities and scaling for general couplings}, 1999 {\it Physica D} {\bf 125} 1

\bibitem{40}
Chiba H, {\it A proof of the Kuramoto conjecture for a bifurcation structure of the infinite-dimensional Kuramoto model}, 2015 {\it Ergodic Theory and Dynamical Systems} {\bf 35} 762

\bibitem{41}
Ott E and Antonsen TM, {\it Low-dimensional behavior of large systems of globally coupled oscillators}, 2008 {\it Chaos} {\bf 18} 037113

\bibitem{42}
Watts DJ and Strogatz SH, {\it Collective dynamics of small-world networks}, 1998 {\it Nature} {\bf 393} 440

\bibitem{43}
Barab\'asi A-L and Albert R, {\it Emergence of scaling in random networks}, 1999 {\it Science} {\bf 286} 509

\bibitem{44}
Rodrigues FA, Peron TK, Ji P and Kurths J, {\it The Kuramoto model in complex networks}, 2016 {\it Phys. Rep.} {\bf 610} 1

\bibitem{45}
Mill\'an AP, Torres JJ and Bianconi G, {\it Explosive higher-order Kuramoto dynamics on simplicial complexes}, 2020 {\it Phys. Rev. Lett.} {\bf 124} 218301

\bibitem{46}
Skardal PS and Arenas A, {\it Higher-order interactions in complex networks of phase oscillators promote abrupt synchronization switching}, 2020 {\it Commun. Phys.} {\bf 3} 218

\bibitem{47}
Yeung MKS and Strogatz SH, {\it Time delay in the Kuramoto model of coupled oscillators}, 1999 {\it Phys. Rev. Lett.} {\bf 82} 648

\bibitem{48}
Tanaka HA, Lichtenberg AJ and Oishi S, {\it First order phase transition resulting from finite inertia in coupled oscillator systems}, 1997 {\it Phys. Rev. Lett.} {\bf 78} 2104

\bibitem{49}
Chandra S, Girvan M and Ott E, {\it Continuous versus discontinuous transition in the D-dimensional generalized Kuramoto model: odd D is different}, 2019 {\it Phys. Rev. X} {\bf 9} 011002

\bibitem{50}
Rohden M, Sorge A, Timme M and Witthaut D, {\it Self-organized synchronization in decentralized power grids}, 2012 {\it Phys. Rev. Lett.} {\bf 109} 064101

\bibitem{51}
Zhang Z, Al-Abri S and Zhang F, {\it A generalized Kuramoto model for opinion dynamics on the unit sphere}, 2025 {\it Automatica} {\bf 171} 111957

\bibitem{52}
Miyato T, L\"owe S, Geiger A and Welling M, {\it Artificial Kuramoto oscillatory neurons}, 2024 arXiv:2410.13821

\bibitem{53}
Zhang X, Boccaletti S, Guan S and Liu Z, {\it Explosive synchronization in adaptive and multilayer networks}, 2015 {\it Phys. Rev. Lett.} {\bf 114} 038701

\bibitem{54}
Tanaka D and Kuramoto Y, {\it Complex Ginzburg-Landau equation with nonlocal coupling}, 2003 {\it Phys. Rev. E} {\bf 68} 026219

\bibitem{55}
Kuramoto Y, {\it Scaling behavior of turbulent oscillators with non-local interaction}, 1995 {\it Prog. Theor. Phys.} {\bf 94} 321

\bibitem{56}
Kuramoto Y and Nakao H, {\it Origin of power-law spatial correlations in distributed oscillators and maps with nonlocal coupling}, 1996 {\it Phys. Rev. Lett.} {\bf 76} 4352

\bibitem{57}
Kuramoto Y, Battogtokh D and Nakao H, {\it Multiaffine chemical turbulence}, 1998 {\it Phys. Rev. Lett.} {\bf 81} 3543

\bibitem{58}
Abrams DM and Strogatz SH, {\it Chimera states for coupled oscillators}, 2004 {\it Phys. Rev. Lett.} {\bf 93} 174102

\bibitem{59}
Shima S and Kuramoto Y, {\it Rotating spiral waves with phase-randomized core in nonlocally coupled oscillators}, 2004 {\it Phys. Rev. E} {\bf 69} 036213

\bibitem{60}
Hagerstrom AM, Murphy TE, Roy R, H\"ovel P, Omelchenko I and Sch\"oll E, {\it Experimental observation of chimeras in coupled-map lattices}, 2012 {\it Nature Physics} {\bf 8} 658

\bibitem{61}
Tinsley MR, Nkomo S and Showalter K, {\it Chimera and phase-cluster states in populations of coupled chemical oscillators}, 2012 {\it Nature Physics} {\bf 8} 662

\bibitem{62}
Martens EA, Thutupalli S, Fourri\`ere A and Hallatschek O, {\it Chimera states in mechanical oscillator networks}, 2013 {\it Proc. Nat. Acad. Sci.} {\bf 110} 10563

\bibitem{63}
Parastesh F, Jafari S, Azarnoush H, Shahriari Z, Wang A, Boccaletti S and Perc M, {\it Chimeras}, 2021 {\it Phys. Rep.} {\bf 898} 1

\bibitem{64}
Nakao H, {\it Phase reduction approach to synchronization of nonlinear oscillators}, 2016 {\it Contemp. Phys.} {\bf 57} 188

\end{thebibliography}
\end{document}